\documentclass[letter,twocolumn]{jpsj3}

\title{
Emergent Electric Polarization by Kondo Effect \\
in a Triangular Triple Quantum Dot
}

\author{Mikito Koga$^1$, Masashige Matsumoto$^2$, and Hiroaki Kusunose$^3$}

\inst{
$^1$Department of Physics, Faculty of Education, Shizuoka University, Shizuoka
422--8529, Japan \\
$^2$Department of Physics, Faculty of Science, Shizuoka University, Shizuoka
422--8529, Japan \\
$^3$Department of Physics, Ehime University, Matsuyama 790-8577, Japan
}

\abst{
A triangular triple quantum dot gives various Kondo effects, such as the emergence of an
electric polarization accompanied by a complete compensation of spin degrees of freedom.
The interplay of spin and charge of electrons in quantum dots is investigated using an
Anderson model with an equilateral triangular spin cluster, in which a single electron state at
one site is hybridized with a conduction band in a lead.
The numerical renormalization group analysis shows how a nonzero electric polarization
develops with local electrons traveling in the loop and how it depends on the hybridization
strength as an experimentally controllable parameter in the Kondo effect.
}

\begin{document}

\maketitle

\newcommand{\ds}{\displaystyle}
\renewcommand{\H}{\mathcal{H}}
\newcommand{\br}{{\mbox{\boldmath$r$}}}
\newcommand{\bR}{{\mbox{\boldmath$R$}}}
\newcommand{\bS}{{\mbox{\boldmath$S$}}}
\newcommand{\bk}{{\mbox{\boldmath$k$}}}
\newcommand{\bH}{{\mbox{\boldmath$H$}}}
\newcommand{\bh}{{\mbox{\boldmath$h$}}}
\newcommand{\bJ}{{\mbox{\boldmath$J$}}}
\newcommand{\bPsi}{{\mbox{\boldmath$\Psi$}}}
\newcommand{\bpsi}{{\mbox{\boldmath$\psi$}}}
\newcommand{\bPhi}{{\mbox{\boldmath$\Phi$}}}
\newcommand{\bd}{{\mbox{\boldmath$d$}}}
\newcommand{\bG}{{\mbox{\boldmath$G$}}}
\newcommand{\om}{{\omega_n}}
\newcommand{\omm}{{\omega_{n'}}}
\newcommand{\omd}{{\omega^2_n}}
\newcommand{\omt}{{\tilde{\omega}_{n}}}
\newcommand{\ommt}{{\tilde{\omega}_{n'}}}
\newcommand{\btau}{{\hat{\tau}}}
\newcommand{\brho}{{\mbox{\boldmath$\rho$}}}
\newcommand{\bsigma}{{\mbox{\boldmath$\sigma$}}}
\newcommand{\bSigma}{{\mbox{\boldmath$\Sigma$}}}
\newcommand{\bI}{{\hat{I}}}
\newcommand{\bt}{{\hat{t}}}
\newcommand{\bq}{{\hat{q}}}
\newcommand{\bLambda}{{\hat{\Lambda}}}
\newcommand{\bDelta}{{\hat{\Delta}}}
\newcommand{\bu}{{\hat{u}}}
\newcommand{\bU}{{\hat{U}}}
\newcommand{\bskp}{{\mbox{\scriptsize\boldmath $k$}}}
\newcommand{\skp}{{\mbox{\scriptsize $k$}}}
\newcommand{\bsrp}{{\mbox{\scriptsize\boldmath $r$}}}
\newcommand{\bsRp}{{\mbox{\scriptsize\boldmath $R$}}}
\newcommand{\bsk}{\bskp}
\newcommand{\sk}{\skp}
\newcommand{\bsr}{\bsrp}
\newcommand{\bsR}{\bsRp}
\newcommand{\ri}{{\rm i}}
\newcommand{\re}{{\rm e}}
\newcommand{\rd}{{\rm d}}
\newcommand{\rM}{{\rm M}}
\newcommand{\rs}{{\rm s}}
\newcommand{\rt}{{\rm t}}

The Kondo effect plays an important role in nanoscale devices, such as quantum dots,
\cite{Kondo05}
atomic contacts,
\cite{Lucignano09,Parks10}
and molecular spins on a metallic surface.
\cite{Komeda11,DiLullo12,Minamitani12}
It is desirable to detect novel quantum phenomena in these contexts that differ from a
conventional case of magnetic impurities embedded in a metal.
In the conventional Kondo effect, spin degrees of freedom of the impurities are coupled to
conduction electrons antiferromagnetically via an effective exchange interaction.
As a consequence of the complete compensation of the local spin moment, the low-temperature
physics is described by the local Fermi-liquid theory.
\cite{Hewson93}
This scenario gives us an essential understanding of the Kondo effect.
However, it is not completed if the impurities have internal degrees of freedom other than spin.
The multichannel Kondo effect is one of the examples associated with the additional degrees of
freedom of an impurity ion.
\cite{Cox98}
The recent development of nanotechnology enables more ideal setups
in artificial atomic devices that could distinguish various types of Kondo effects in addition to the
conventional single-channel and two-channel Kondo effects.
\cite{Potok07}
\par

The advantage of nanoscale devices for the Kondo effect is their controllability of various
parameters, for instance, the hybridization strength of a localized electron state with a conduction
band by the fine tuning of the nanocontact between an artificial atom and metallic leads.
It is also expected that the assemblage of artificial atoms will produce various structural properties
analogous to real atoms or molecules.
In the triangular triple quantum dot (TTQD), the electrons traveling around in the loop
cause interdot spin correlations owing to the Coulomb repulsion on each dot.
The three-dot configuration of TTQD can be regarded as a variant of a threefold orbitally
degenerate atomic state.
\cite{Koga98}
The simplest description of TTQD is given by a Hubbard model that is reduced to a Heisenberg
spin model with antiferromagnetic interaction in the half-filled case with a strong repulsion.
Bulaevskii {\it et al.} argued a novel mechanism of a nonzero electric 
polarization, which is induced by magnetic ordering in a Mott insulator or by a magnetic field in a
triangular lattice system.
\cite{Bulaevskii08,Matsumoto12,Kamiya12}
In the strong Coulomb coupling limit, a single electron is completely localized on each site.
In triangular spin networks, each-site electron occupation number can deviate from one owing to
electron hopping in the loop.
This is the key to an emergent electric polarization that depends on the spin structure.
This is also possible for the Kondo effect in TTQD and is controllable by the fine tuning of a
point contact between the apex of TTQD and the lead.
It is the so-called multiferroic device in nanoscale, contrary to the macroscopic scale of
electric polarization accompanied by a magnetic phase transition in multiferroic materials.
\cite{Wadati12}
\par

Various types of Kondo effects in TTQD have been studied theoretically on the basis of an
extended Anderson model or Kondo model for different configurations of TTQD and leads.
\cite{Oguri07,Zitko08,Mitchell09,Vernek09,Oguri11}
Extensive studies have focused on the variation of ground states and the conductance through the
leads, which depend sensitively on the geometric structure, electron filling, and gate voltage
potential of TTQD.
However, not much attention has been paid to a charge redistribution that is another intriguing
feature of TTQD with a loop.
In the previous studies, the electron hopping parameter of TTQD is taken to be extremely 
small or fixed as an energy unit.
\par

In this letter, we present an idea of emergent electric polarization in a spin-controlled device.
The induced electric dipole moment is small in nanoscale but can be controlled experimentally by
the Kondo effect.
For the isolated TTQD, we assume that three sites are completely identical, reflecting an
equilateral triangular geometry.
Recently, such a geometric TTQD device has been targeted for experimental development.
\cite{Amaha08,Amaha09}
For quantitative analysis, we use Wilson's numerical renormalization group (NRG) method.
\cite{Wilson75,Krishnamurthy80}
This is a powerful and unbiased tool since a possibly large number of states are taken into account
systematically in the calculation with the logarithmic discretization of a conduction band in
the renormalization procedure.
In particular, the effectiveness of this tool is more evident in the investigation of the low-temperature
physics associated with the very complex structure of a quantum-dot cluster.
\par

\begin{figure}
\begin{center}
\includegraphics[width=7cm,clip]{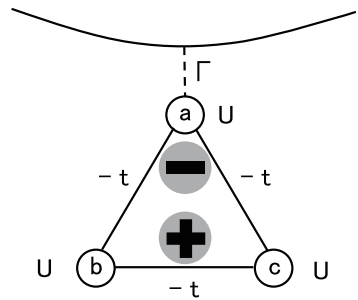}
\end{center}
\caption{
Schematic illustration of TTQD.
The apex (labelled $a$) of the triangular cluster is coupled to the metallic lead through the
hybridization (represented by the dotted line) between the localized and conduction electron
states.
The electric polarization is induced by the Kondo effect at low temperatures.
}
\label{fig:1}
\end{figure}

Throughout this letter, we consider a half-filled case for TTQD in the strong Coulomb coupling
limit.
Let us begin with an isolated three-site Hubbard-type model Hamiltonian,
\begin{align}
& H_{\rm dot} = -t \sum_{i \ne j} \sum_{\sigma} ( d_{i \sigma}^{\dagger} d_{j \sigma}
+ d_{j \sigma}^{\dagger} d_{i \sigma} ) \cr
&~~~~~~
+ \varepsilon_{\rm d} \sum_i n_i + U \sum_i n_{i \uparrow} n_{i \downarrow},
\label{eqn:Hdot}
\end{align}
where the three sites (labelled $i,j = a,b,c$) consist of an equilateral triangular cluster.
On the $i$-th site, $d_{i \sigma}^{\dagger}$ ($d_{i \sigma}$) is the creation (annihilation)
operator of an electron with spin $\sigma$ ($= \uparrow, \downarrow$) and
$n_i = n_{i \uparrow} + n_{i \downarrow}$
($n_{i \sigma} = d_{i \sigma}^{\dagger} d_{i \sigma}$) is the number operator.
The first term in eq.~(\ref{eqn:Hdot}) represents electron hopping between the
nearest-neighbor sites with a positive parameter $t$.
In the other terms, $\varepsilon_{\rm d}$ is the energy of a localized orbital,
and $U$ is the on-site Coulomb coupling.
In the following analysis, we restrict ourselves to the symmetric condition
($\varepsilon_{\rm d} = - U / 2 < 0$) for each site.
Since a single electron is almost localized on each site for $t / U \ll 1$, 
the low-energy subspace is described by the on-site spin $1/2$ operators $\bS_i$.
The three-site loop geometry indicates the appearance of a physical quantity associated with the
lifting of spin degeneracy via the third-order perturbation of $t / U$.
According to Bulaevskii {\it et al.},
\cite{Bulaevskii08}
the charge operator is given by
\begin{align}
\hat{n}_a = 1 + 8 \left( \frac{t}{U} \right)^3 [ \bS_a \cdot ( \bS_b + \bS_c ) - 2 \bS_b \cdot \bS_c ]
\label{eqn:na}
\end{align}
on the $a$ site ($\hat{n}_b$ and $\hat{n}_c$ are given by the cyclic permutation of the three
indices).
Owing to the equilateral triangular symmetry, the ground state is fourfold degenerate.
\cite{Mitchell09}
We represent the ground-state wave functions for the total spin with $S = 1/2$ and
$S_z = 1/2$ as
\begin{align}
& | \phi_{{\rm g} +} \rangle = \frac{1}{\sqrt{2}} d_{a \uparrow}^{\dagger}
( d_{b \uparrow}^{\dagger} d_{c \downarrow}^{\dagger}
- d_{b \downarrow}^{\dagger} d_{c \uparrow}^{\dagger} ) | 0 \rangle,
\label{eqn:phig+} \\
& | \phi_{{\rm g} -} \rangle = \frac{1}{\sqrt{6}} [ d_{a \uparrow}^{\dagger}
( d_{b \uparrow}^{\dagger} d_{c \downarrow}^{\dagger}
+ d_{b \downarrow}^{\dagger} d_{c \uparrow}^{\dagger} )
- 2 d_{a \downarrow}^{\dagger} d_{b \uparrow}^{\dagger} d_{c \uparrow}^{\dagger} ] | 0 \rangle,
\label{eqn:phig-}
\end{align}
where $| 0 \rangle$ is a vacuum state.
As the time reversal partners, the $S_z = - 1/2$ states are given by interchanging the spin-up and
spin-down in the wave functions.    
Using eq.~(\ref{eqn:na}), we obtain
\begin{align}
& \langle \phi_{{\rm g} \pm} | \hat{n}_a | \phi_{{\rm g} \pm} \rangle
= 1 + 8 \left( \frac{t}{U} \right)^3 \left( \pm \frac{3}{2} \right), \\
& \langle \phi_{{\rm g} \pm} | \hat{n}_b | \phi_{{\rm g} \pm} \rangle
= \langle \phi_{{\rm g} \pm} | \hat{n}_c  | \phi_{{\rm g} \pm} \rangle
= 1 + 8 \left( \frac{t}{U} \right)^3 \left( \mp \frac{3}{4} \right).
\label{eqn:ng}
\end{align}
As long as the ground state maintains the degeneracy, the expectation value of the electron
occupation on each site is given by $\langle \hat{n}_i \rangle = 1$.
Equation~(\ref{eqn:na}) implies that a charge redistribution occurs if the three spins become
inequivalent.
In the present study, we consider the point contact between a conduction electron
system and a single site in the triangular cluster, for instance, $\bS_a$ (see Fig.~\ref{fig:1}).
When $\bS_a$ is quenched by the Kondo effect, eq.~(\ref{eqn:na}) gives
\begin{align}
\hat{n}_a = 1 + 8 \left( \frac{t}{U} \right)^3 (- 2 \bS_b \cdot \bS_c).
\label{eqn:naK}
\end{align}
Since the electron hopping between the $b$ and $c$ sites stabilizes a local singlet with
$\langle \bS_b \cdot \bS_c \rangle < 0$, the charge redistribution $\langle \hat{n}_a \rangle > 1$
results in the induction of a nonzero electric dipole moment in TTQD.
It is noted that the total spin is completely quenched by the Kondo effect and the pure electric polarization is realized in TTQD.
\par

We analyze quantitatively how the electric polarization is induced by the Kondo effect in TTQD
using the following Anderson model Hamiltonian:
\cite{Mitchell09}
\begin{align}
& H = H_{\rm lead} + H_{\rm dot} + H_{\rm mix}, 
\label{eqn:Ham} \\
&~~
H_{\rm lead} = \sum_{\bsk \sigma} \varepsilon_{\bsk} c_{\bsk \sigma}^{\dagger} c_{\bsk \sigma}, \\
&~~
H_{\rm mix} = \sum_{\bsk \sigma} ( v_{\bsk} d_{a \sigma}^{\dagger} c_{\bsk \sigma}
+ v_{\bsk}^* c_{\bsk \sigma}^{\dagger} d_{a \sigma} ).
\end{align}
In eq.~(\ref{eqn:Ham}), the first term represents the kinetic energy $\varepsilon_{\bsk}$ of
electrons with the wave vector $\bk$ in the lead for which $c_{\bsk}^{\dagger}$ ($c_{\bsk}$) is
a creation (annihilation) operator.
The second term is given by eq.~(\ref{eqn:Hdot}).
The last term represents the electron transfer between the $a$ site and the lead.
Here, the hybridization strength $v_{\bsk}$ is considered to be a constant and related to the level
broadening of $\Gamma \equiv \pi \rho |v_{\bsk}|^2$ ($\rho$ is the density of states at the Fermi
energy).
Following Wilson, we transform eq.~(\ref{eqn:Ham}) to the hopping-type Hamiltonian for NRG 
calculation.
\cite{Wilson75,Krishnamurthy80}
The numerical analysis is carried out by the recursion relation of the form
\begin{align}
& H_{N+1} = \Lambda^{1/2} H_N + \sum_{\sigma} \xi_N ( f_{N \sigma}^{\dagger} f_{N+1,\sigma}
+ f_{N+1,\sigma}^{\dagger} f_{N \sigma} ), \\
&~~ H_0 = \Lambda^{-1/2}
\left[ -\tilde{t} \sum_{i \ne j} \sum_{\sigma} ( d_{i \sigma}^{\dagger} d_{j \sigma}
+ d_{j \sigma}^{\dagger} d_{i \sigma} ) \right. \cr
&~~\left.
+ \frac{\tilde{U}}{2} \sum_i ( n_i - 1 )^2
+ \tilde{\Gamma}^{1/2} \sum_{\sigma} ( f_{0 \sigma}^{\dagger} d_{a \sigma}
+ d_{a \sigma}^{\dagger} f_{0 \sigma} ) \right],
\end{align}
\begin{align}
\xi_N = (1 - \Lambda^{-N-1})(1 - \Lambda^{-2N-1})^{-1/2} (1 - \Lambda^{-2N-3})^{-1/2},
\end{align}
where $\Lambda$ is a parameter for the logarithmic discretization of the conduction band, and the
parameters are scaled as
\begin{align}
& \tilde{t} = \frac{2}{1 + \Lambda^{-1}} \frac{t}{D},~~
\tilde{U} = \frac{2}{1 + \Lambda^{-1}} \frac{U}{D}, \cr
& \tilde{\Gamma} = \left( \frac{2}{1 + \Lambda^{-1}} \right)^2 \frac{2 \Gamma A_{\Lambda}}{\pi D}~~
\left( A_\Lambda = \frac{1}{2} \frac{1 + \Lambda^{-1}}{1 - \Lambda^{-1}} \ln \Lambda \right).
\end{align}
The energy is normalized by the half width of the conduction band $D$ and $A_{\Lambda}$ is
a correction parameter related to the continuum limit $\Lambda \rightarrow 1$.
Throughout the NRG calculation, we consider $\Lambda = 3$ and maintain about $2000$
lowest-lying states at each renormalization step.
The electron occupation at each site is calculated by
\begin{align}
\langle n_i \rangle = \frac{{\rm Tr}~n_i \exp (- \bar{\beta} H_N)}{{\rm Tr}~\exp (- \bar{\beta} H_N)}~~
(i = a,b,c),
\end{align}
where $\bar{\beta}$ ($\sim 1$) is related to the physical temperature
$T / D = [(1 + \Lambda^{-1}) / 2] \Lambda^{-(N-1)/2} / \bar{\beta}$.
By these quantities, we evaluate the average electron number and electric polarization of TTQD as
$\langle n_{\rm dot} \rangle \equiv (1/3) \sum_{i = a, b, c} \langle n_i \rangle$
and
\begin{align}
\delta n \equiv \langle n_a \rangle - \langle n_{\rm dot} \rangle
= \frac{1}{3} ( 2 \langle n_a \rangle -  \langle n_b \rangle - \langle n_c \rangle ),
\end{align}
respectively.
\par

We examine the Kondo effect in TTQD for $U / D \sim 1$ and $\pi \Gamma / U \ll 1$, where a
local spin moment is well developed at the $a$ site.
For $\Gamma  =0$, TTQD is isolated from the conduction electron system and the ground
state is fourfold degenerate as mentioned above.
This degeneracy is lifted by a finite $\Gamma$ since the effective antiferromagnetic exchange
between the local spin $\bS_a$ and conduction electrons leads to the Kondo singlet at low
temperatures corresponding to a large number of renormalization steps.
The low-temperature physics is described by the local Fermi-liquid theory.
\cite{Hewson93}
The remaining two local spins, $\bS_b$ and $\bS_c$, are combined to be a singlet by the
superexchange coupling.
\cite{Mitchell09}
As a result, the electric polarization emerges as a whole.
\par
\begin{figure}
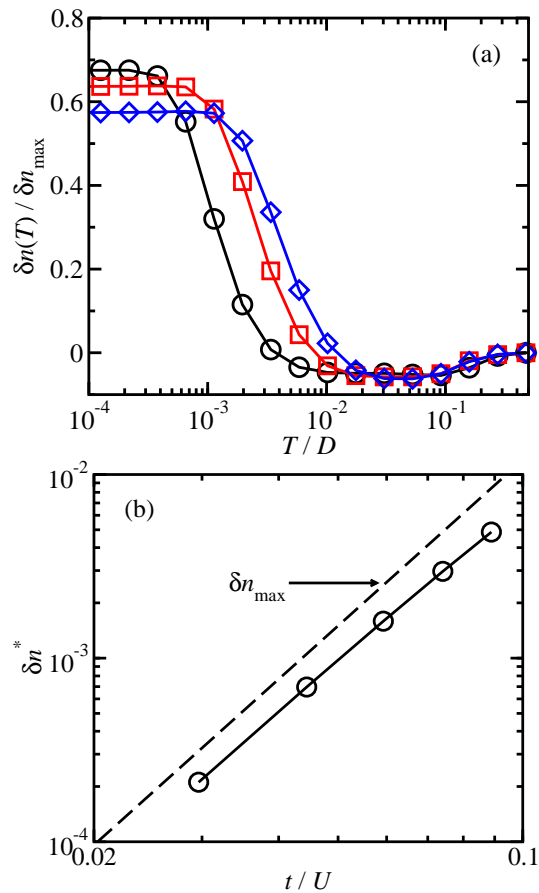

\begin{center}
\includegraphics[width=7cm,clip]{fig2a.eps}
\includegraphics[width=7cm,clip]{fig2b.eps}
\end{center}
\caption{
(Color online)
(a)~Temperature dependence of the induced electric polarization $\delta n$ for $t / U = 0.03$
(circles),
$0.06$ (squares), and $0.09$ (diamonds).
The unit of the electric polarization is defined as $\delta n_{\rm max} = 12 (t  /U)^3$.
(b)~$t / U$ dependence of the saturated electric polarization $\delta n^*$.
Both $U / D = 0.9$ and $\Gamma / U = 0.0946$ are fixed here.
$\delta n^*$ obeys the power law of $(t / U)^3$ approximately and lies under the
$\delta n_{\rm max}$ line for various $\Gamma$ values.
The deviation from the $(t / U)^3$ dependence is shown in Fig.~3.
}
\label{fig:2}
\end{figure}

In Fig.~\ref{fig:2}(a), we show the temperature dependence of $\delta n$.
$\delta n$ reaches a constant ($\equiv \delta n^*$) at low temperatures where the local state in
TTQD can be regarded as a singlet.
As shown in Fig.~\ref{fig:2}(b), we find that the saturated electric polarization $\delta n^*$ is smaller
than the maximum value given by
\begin{align}
\delta n^* < \delta n_{\max} \equiv 12 \left( \frac{t}{U} \right)^3.
\end{align}
Here, the right-hand side is obtained by taking $\langle \bS_b \cdot \bS_c \rangle = - 3/4$ in
eq.~(\ref{eqn:naK}) that corresponds to a local singlet in the $b$-$c$ bond.
It is convenient to normalize $\delta n$ by $\delta n_{\rm max}$ to examine how the electric
polarization depends on the Kondo effect for a fixed $\Gamma / U$.
In Figs.~\ref{fig:2}(a) and \ref{fig:2}(b), $\delta n^* / \delta n_{\rm max}$ increases as $t / U$
decreases.
On the other hand, the upturn of $\delta n \simeq 0 \rightarrow \delta n^*$ begins at the higher
temperature for the larger $t / U$.
Thus, the electric polarization by the Kondo effect in TTQD is sensitive to $t / U$.
In the high-temperature region ($10^{-2} < T / D < 10^{-1}$), $\delta n \simeq 0$ indicates that the
ground state maintains the fourfold degeneracy of TTQD, which is more robust for the smaller
$t / U$.
This is a characteristic feature of competing phenomena.
\cite{Jones87}
In our case, the smaller $t$ favors the isolated spins at high temperatures, while the Kondo effect
tends to form the local spin singlet with conduction electrons at low temperatures.
In view of eqs.~(\ref{eqn:phig+})--(\ref{eqn:ng}), the degeneracy of
$| \phi_{{\rm g} +} \rangle$ and $| \phi_{{\rm g} -} \rangle$ is lifted by different Kondo couplings
with the conduction electrons that break the equivalency of the three spins in TTQD.
Thus, the development of electric polarization with the decrease in temperature indicates the
crossover from the Heisenberg spin cluster on the triangle to the Kondo singlet at the $a$ site
plus the local singlet on the $b$-$c$ bond.
\par
\begin{figure}
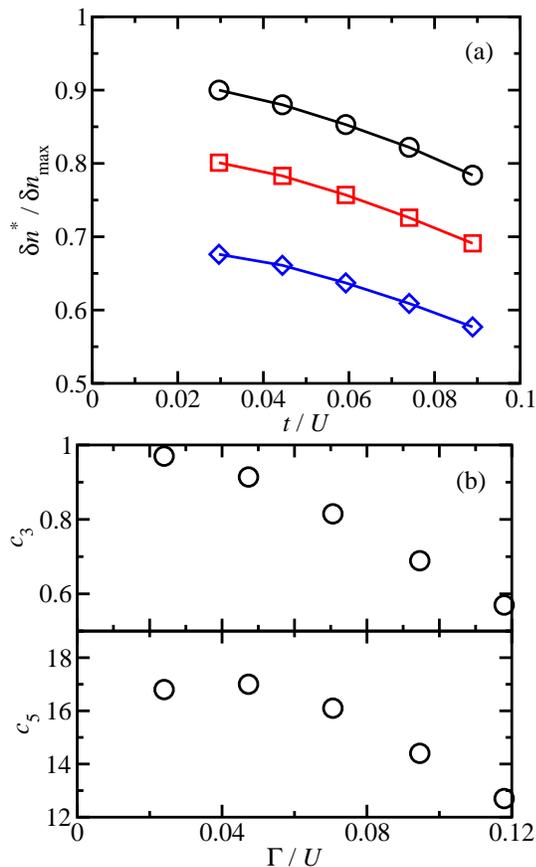

\begin{center}
\includegraphics[width=7cm,clip]{fig3a.eps}
\includegraphics[width=7cm,clip]{fig3b.eps}
\end{center}
\caption{
(Color online)
Saturated electric polarization $\delta n^*$ at low temperatures.
$U / D = 0.9$ is fixed here.
(a)~$t / U$ dependence for $\Gamma / U = 0.0473$ (circles), $0.0706$ (squares), and $0.0946$
(diamonds), normalized by $\delta n_{\rm max} = 12 (t  /U)^3$.
(b)~Evaluated $c_3$ and $c_5$ as a function of $\Gamma / U$ in the fitting of
$\delta n^* = 12 [c_3 (t/U)^3 - c_5 (t / U)^5]$.
}
\label{fig:3}
\end{figure}

The NRG result shows that the electric polarization induced in TTQD is also very sensitive to
$\Gamma / U$.
The data are given in Fig.~\ref{fig:3}.
Regardless of the $\Gamma / U$ values, $\delta n^*$ is proportional to $( t / U )^3$ approximately.
This means that eq.~(\ref{eqn:na}) holds for $\Gamma > 0$, and we use
$\delta n_{\rm max}$ for an appropriate unit to extract the contribution of the Kondo effect to the
electric polarization.
In Fig.~\ref{fig:3}(a), $\delta n^* / \delta n_{\rm max}$ is more increased with the smaller $t / U$.
Considering a higher order correction with respect to $t / U$, we find $\delta n^*$ fitted as
\begin{align}
\delta n^* = 12 \left[ c_3 \left( \frac{t}{U} \right)^3 - c_5 \left( \frac{t}{U} \right)^5 \right],
\end{align}
where the evaluated $c_3$ and $c_5$ are plotted as a function of $\Gamma / U$ in
Fig.~\ref{fig:3}(b).
The small $t / U$ limit $c_3$ of $\delta n^* / \delta n_{\rm max}$ increases for the smaller
$\Gamma / U$, and $c_3$ approaches unity when $\Gamma / U$ is close to zero.
The indication of $\delta n^* / \delta n_{\rm max} \rightarrow 1$ also appears
in the $\Gamma / U$ dependence of $c_5$ that reaches the peak at the small $\Gamma / U$.
The strength of $\Gamma$ can be controlled experimentally by changing the point contact
between the edge of the lead and the apex of TTQD.
For practical use, $\delta n^*$ should be large at an appropriate parameter range.
Thus, the optimization of $\Gamma$ is important.
The emergence of an electric polarization is also controllable by an applied magnetic field, which
suppresses the Kondo effect.
\par

Finally, we would like to mention a connection between the present TTQD and an atomic structure
with orbital degeneracy.
The $a$ site is regarded as a delocalized orbital since it is hybridized with the conduction band.
This distinct site is coupled to the other sites through the electron transfer within TTQD.
It is different from a real atom in which degenerate orbitals are correlated by intra-atomic
interactions.
The variation of the Kondo effect arises depending on the degrees of orbital localization,
\cite{Koga99,Koga07}
which has been investigated extensively in heavy fermion systems,
\cite{Cox98}
and many analogies with such orbital dynamics are applicable to nanoscale phenomena in
multiple quantum dot systems.
\par

In conclusion, we showed how an electric polarization is induced by the Kondo effect in TTQD
that breaks the equivalency of the three spins on the triangle.
The emergent electric dipole moment can be enhanced at lower temperatures by the weaker
hybridization of TTQD with the lead,  which is expected to be experimentally controllable.
\par

\acknowledgments
We are indebted to T. Kawae for sending us the update of experimental development in
nano-Kondo physics.
This work is supported by a Grant-in-Aid for Scientific Research~C (No. 23540390) from the Japan
Society for the Promotion of Science.
One of the authors (H.K.) is supported by a Grant-in-Aid for Scientific Research on Innovative
Areas "Heavy Electrons" (No. 20102008) from The Ministry of Education, Culture, Sports, Science,
and Technology, Japan.


\end{document}